\documentclass[twocolumn,prb,amsmath,superscriptaddress,citeautoscript,floatfix,showpacs]{revtex4}
\usepackage{graphicx}

\usepackage{amssymb}
\usepackage{color}
\usepackage{epsfig}

\def\b0{{\bf{0}}}

\begin{document}

\title{Efficiency and dissipation in   a two-terminal thermoelectric junction, \\ emphasizing small dissipation \\ }

\author{O. Entin-Wohlman}
\affiliation{
Raymond and Beverly Sackler School of Physics and Astronomy, Tel Aviv University, Tel Aviv 69978, Israel
}\affiliation{Department of Physics and the Ilse Katz Center for Meso-
  and Nano-Scale Science and Technology, Ben Gurion University, Beer
  Sheva 84105, Israel}

\author{J-H. Jiang}

\affiliation{Department of Condensed matter Physics, Weizmann Institute of
  Science, Rehovot 76100, Israel}
\affiliation{Present address: Department of Physics, University of Toronto, 60 St. George St., Toronto, Canada ON M5S 1A7}

\author{Y. Imry}
\affiliation{Department of Condensed matter Physics, Weizmann Institute of
  Science, Rehovot 76100, Israel}

\date{\today}

\begin{abstract}
The efficiency and cooling power of a two-terminal thermoelectric refrigerator are analyzed near the 
limit of vanishing dissipation (ideal system), where the optimal efficiency is the Carnot one, but the cooling power then unfortunately vanishes. This limit, where transport occurs only via a single sharp electronic energy, has been referred to as ``strong coupling" or ``the best thermoelectric". It follows however, that  ``parasitic" effects that make the system deviate from the ideal limit, and reduce the efficiency from the Carnot limit, are crucial for the usefulness of the device. Among these parasitics, there are: parallel phonon conduction, finite width of the electrons' transport band and more than a single energy transport channel. In terms of a small parameter characterizing the deviation from the ideal limit, the efficiency and power grow linearly, and the dissipation {\em quadratically}. The results are generalized to the case of broken time-reversal symmetry, and the major nontrivial changes are discussed. Finally, the recent universal relation between the thermopower and the asymmetry of the dissipation between the two terminals is briefly discussed, including the small dissipation limit.
\end{abstract}

\maketitle

\section{Introduction}

There has been major interest in thermoelectric energy conversion \cite{book,Kedem,Rutten,we,Cleuren,JHJPAP,NJP,Mahan}. An important application  is converting wasted thermal energy into, say, useful and storable electrical one. Also, deriving  electricity from the more easily 
available  thermal energy in remote environments is often relevant. \cite{Rutten,Cleuren} Thermoelectric cooling is a real possibility. 
Obviously, there exist rigid thermodynamic limitations to such processes. However, the efficiency of existing devices is severely further limited 
by material and device properties, \cite{gordon} so that getting a sizable fraction of the thermodynamic Carnot efficiency (especially with a reasonable power) is still an unattained goal.

A key idea in this connection is that of trying to obtain a ``strongly-coupled" thermoelectric,   \cite{Kedem}
in which there would be a well-defined ratio between electrical and thermal current. Mahan and Sofo \cite{Mahan}  suggested a way to achieve that, for electrons, by funneling the electrical transport to a very narrow energy-band. The deviation from the ideal Carnot efficiency would then be determined by
``parasitic" effects, such as the small width of the above-mentioned  electronic band and the non-electronic thermal conductivity.

For definiteness, we shall consider here cooling in a two-terminal electronic setup, whose 
terminals are kept at different temperatures, $T_{L}$ and $T_{R}$,  and  different chemical potentials, $\mu_{L}$ and $\mu_{R}$.
The total entropy production of the electrons in such a device is
\begin{align}
\dot{S}^{}_{\rm tot}&=\frac{\dot{E}^{}_{L}-\mu^{}_{L}\dot{N}^{}_{L}}{T^{}_{L}}+
\frac{\dot{E}^{}_{R}-\mu^{}_{R}\dot{N}^{}_{R}}{T^{}_{R}}\equiv\frac{\dot{Q}^{}_{L}}{T^{}_{L}}+
\frac{\dot{Q}^{}_{R}}{T^{}_{R}}\ .\label{sdot}
\end{align}
Here, $\dot{E}_{L,R}$ is the rate of energy change in the $L,R$  reservoir,  $\dot{N}_{L,R}$ is {\em minus} the particle current leaving that reservoir, and  
$\dot{Q}_{L,R}$ is   the  heat current entering  it. 
Current conservation implies $\dot{N}_{L}+\dot{N}_{R}=0$, and energy conservation requires 
$\dot{E}_{L}+\dot{E}_{R}=0$. \cite{COM2}
Consequently, 
%
%
 \begin{align}
\dot{S}^{}_{\rm tot}&=
J^{Q}_{L}\Big (\frac{1}{T^{}_{R}}-\frac{1}{T^{}_{L}}\Big )+\frac{J^{}_{L}}{T^{}_{R}}(\mu^{}_{L}-\mu^{}_{R} )\ ,\label{sdot1}
\end{align}
where $J^{Q}_{L}\equiv -\dot{E}_{L}+\mu_{L}\dot{N}_{L}$ is the heat current and $J^{}_{L}\equiv -\dot{N}_{L}$ is the particle current leaving the left reservoir.

Assuming  for concreteness that the left terminal is to be cooled, then $T_{L}-T_{R}<0$
and
the working conditions for
a thermoelectric refrigerator configuration are
positiveness of the used  power  $W$
\begin{align}
W=\frac{\mu^{}_{L}-\mu^{}_{R}}{e}eJ^{}_{L}\equiv VJ^{}_{L}>0\ ,
\end{align}
($e$ being the charge of the carriers and $V$ is the voltage drop) and positiveness of the cooling power $P$, 
\begin{align}
P\equiv J^{Q}_{L}>0\ .
\end{align}
The efficiency $\eta $ of the cooling process (sometimes called ``coefficient of performance") is  defined as the ratio between the cooling power and the used one; it is advantageous to express it in terms of the entropy production   [see Eq. (\ref{sdot1})]          i.e., the dissipation \cite{Cleuren}
\begin{align}
\eta=\frac{P}{W}=\eta^{}_{\rm C}\Big (1-\frac{T^{}_{R}\dot{S}^{}_{\rm tot}}{W}\Big )\ .\label{ef}
\end{align}
Here $\eta_{\rm C}$ is  the Carnot efficiency for cooling, $T_{\rm cold}/(T_{\rm hot}-T_{\rm cold}$;  in our setup it is given by
\begin{align}
\eta^{}_{\rm C}=\frac{T^{}_{L}}{T^{}_{R}-T^{}_{L}}\ .
\end{align}
Several physical remarks on the dissipation are called for here.
The total entropy of the world must not decrease. Usually, for example, the ohmic part of the dissipation occurs in the resistor connecting the two reservoirs. In the two-terminal Landauer-type  formulation used here, there is no reservoir taking the heat off that resistor. The dissipation occurs in the two reservoirs $L$ and $R$: \cite{mybook} the charge carriers exiting or entering these reservoirs are supposed to equilibrate (usually by the unmentioned electron-electron, or electron-phonon interactions) in the huge reservoirs. The interesting question of how the dissipation is divided between the two reservoirs has been recently introduced and discussed in Ref. ~\onlinecite{JCC}. 
It is found that the dissipation occurs symmetrically in the two reservoirs 
only in the limit of full electron-hole symmetry (i.e.,  when the energy-dependent conductivity or  transmission of the junction
is symmetric around the average chemical potential) and then the thermopower vanishes. The larger is this asymmetry (and the thermopower) the more asymmetric is the distribution of the dissipation  between the two reservoirs. Thus the  thermopower and the asymmetry of the dissipation  are monotonic functions of one another!
We will come back to this interesting question in Sec.  \ref{dissa}.

Equation  (\ref{ef})
suggests that a vanishing dissipation can lead to a full Carnot efficiency, $\eta=\eta^{}_{\rm C}$.
This is indeed true, but, as is well-known, in that limit  
{\em both the used and the cooling power will also vanish}. This happens for a junction obeying time-reversal symmetry; breaking it modifies this conclusion. \cite{Benenti}
One might be tempted to consider the possibility that the two terms forming the dissipation, Eq. (\ref{sdot1}), 
cancel one another while each remains finite. This might have been possible, since the thermal
part of the dissipation, $J^{Q}_{L}(T^{}_{L}-T^{}_{R})/T^{}_{L}$,  
is in fact negative (the cooling heat current flows against the temperature drop). However, the above is only apparent: at  the zero-dissipation limit all currents and input/output powers vanish. 
One might have hoped to overcome this caveat by breaking time-reversal symmetry; however, at least in a Landauer-type formulation, it has been shown \cite{Seifert} that there are in fact strong limitations (beyond the Carnot one) on $\eta$. Nevertheless, if one considers, as we shall do here, a system very close (characterized by a small parameter $\zeta$) to  the ideal limit of Carnot efficiency, both the used and the cooling powers will vanish proportional to $\zeta$, while the dissipation will vanish like $\zeta^2$.

To appreciate this point, it is instructive
to relate the dissipation to the properties of the Onsager matrix ${\cal M}$, which connects the currents and the thermodynamic  ``driving" forces (sometimes called affinities), 
\begin{align}
\left [\begin{array}{c}eJ^{}_{L} \\ J_{L}^{Q}\end{array}\right ]={\cal M}
\left [\begin{array}{c}(\mu^{}_{L}-\mu_{R}^{})/(eT^{}_{R}) \\ (1/T^{}_{R})-(1/T^{}_{L})\end{array}\right ]\ .\label{onsager}
\end{align}
In the linear-response regime ${\cal M}$ does not depend on the driving  forces and is determined by the equilibrium properties of the setup.
(For a system invariant to time-reversal the matrix ${\cal M}$ is symmetric.)
The dissipation, Eq. (\ref{sdot1}), can be written in the form
\begin{align}
T^{}_{R}\dot{S}^{}_{\rm tot}=\left [\begin{array}{cc}\frac{\mu^{}_{L}-\mu_{R}^{}}{e} &\ \  -\eta^{-1}_{\rm C}\end{array}\right ]\frac{{\cal M}}{T^{}_{R}}
\left [\begin{array}{c}\frac{\mu^{}_{L}-\mu_{R}^{}}{e} \\ \\ -\eta^{-1}_{\rm C}\end{array}\right ]\ .
\end{align}
As the Onsager matrix ${\cal M}$ is positive definite, it follows that the dissipation vanishes when the lowest eigenvalue of ${\cal M}$ is zero.
A 2$\times$2 symmetric matrix whose determinant is zero is necessarily of the form 
\begin{align}
{\cal M}\propto \left [\begin{array}{cc}1&\ \ \alpha \\ \alpha  &\ \ \alpha^{2}\end{array}\right ]\ .        
\end{align}
Such a matrix implies that the charge and the heat currents are proportional to one another. This was termed by Kedem and Caplan \cite{Kedem}  ``strong coupling".  The strong-coupling  configuration is the limit of the  ``best thermoelectric" 
of Mahan and Sofo.  \cite{Mahan}
Thus, if one could achieve the ideal limit of Refs. ~\onlinecite{Kedem} and ~\onlinecite{Mahan}, the ``driving force"  vector
$
\{(\mu^{}_{L}-\mu_{R}^{})/e , -\eta^{-1}_{\rm C}\}$
would be proportional to $\{-\alpha, 1\}$. One would then get the 
Carnot efficiency, but both currents would vanish, along 
with {\em zero power in the linear-response regime.}

Mahan and Sofo \cite{Mahan} achieved the strong-coupling configuration by channeling the electronic transport into an infinitely-narrow band. There, (see also Ref.~ \onlinecite{JHJPAP}) the Onsager matrix (again limited to electronic processes) takes the form
\begin{align}
\frac{\cal M}{T}=G\left [\begin{array}{cc}1&\ \ \ \   E/e\\
E/e&\ \ \ \ \ \  E^2/e^{2}_{} \end{array}\right ]\ ,\label{JHJM}
\end{align}
where $T$ is the common temperature of the setup, 
$G$ is the conductance,  and $E$ is essentially the {\em unique, well-defined,} energy transferred by each charge carrier between the two terminals. %
In real life there are always  ``parasitic" effects, such as phonon heat 
conductivity, transport in a finite-width energy band, more energy channels, {\it etc.}
As we have now just discovered, such ``parasitics" can be {\em crucial for the operation of thermoelectric heat conversion!}
One is thus led to consider what these parasitics do to the dissipation, to the power, and to the efficiency.

From this line of reasoning it follows that we need to consider the situation when the (nonnegative) determinant of the Onsager matrix is positive, but still very small. \cite{COM3}
In the configuration of Refs. ~\onlinecite{JHJPAP} and ~\onlinecite{NJP}
the two-terminal setup is further connected to a  non-electronic {\em thermal} terminal, which exchanges energy with the electronic system.
The Onsager matrix becomes a 3$\times$3 one, ${\cal M}_{3}$, which is still singular (as the electronic transport occurs via a single channel). The 2$\times$2 Onsager matrix obtained from it when the energy flow between the electronic junction and the thermal contact is blocked is singular as well.
However, adding to the picture the phonon heat currents and the corresponding conductances, 
and the various transport coefficients due to elastic and inelastic (electronic) processes
 \cite{NJP} makes the determinant of ${\cal M}_{3}$, and consequently also that of the 2$\times$2 one obtained from it, finite.

We show below that when time-reversal symmetry holds,  the effect of all these parasitic processes
on the maximal efficiency and the cooling and used powers can be described by a single (small) parameter--the deviation of ${\rm det}[{\cal M}]$ from zero. Moreover, the functional dependence of the maximal efficiency and the dissipation on that parameter are different: the total dissipation is affected far less than the efficiency.

In Sec. \ref{calc} we present the details of the calculations. 
The analysis of the optimum driving-force  vectors and the eigenvectors of the Onsager matrix is presented in section \ref{vec}. 
The question of the symmetry of the dissipation (its distribution between the two reservoirs) and its relation to the thermopower \cite{JCC} is further discussed in Sec.  \ref{dissa}, together with our conclusions. We display the general results, when time-reversal symmetry may not hold. It is straightforward to have in mind the more usual case, when it holds and the discussion is simpler.

\vspace{1cm}

\section{Calculations and results}
\label{calc}

As is explained above, the dissipation can vanish when the determinant of the Onsager matrix ${\cal M}$
is zero. In order to study the efficiency and the optimal driving forces at small dissipation, we introduce a  parameter which ``measures" the distance of the determinant from the strong-coupling limit condition
\begin{align}
\zeta&=\sqrt{1-\frac{{\cal M}^{}_{12}{\cal M}^{}_{21}}{{\cal M}^{}_{11}{\cal M}^{}_{22}}}\ ,\ \ \zeta\geq 0\ . 
\end{align}
When  time-reversal symmetry holds the Onsager matrix is symmetric, and $\zeta=(1+(zT))^{-1/2}$, where $zT$ is the figure of merit. In that case $0\leq\zeta\leq 1$. These bounds on $\zeta$ persist also when ${\cal M}$ is not symmetric, with ${\cal M}^{}_{12}/{\cal M}^{}_{21}>0$. However, when that ratio is negative 
(the example considered below will be  ${\cal M}_{21}<0$)  then $\zeta$  exceeds 1. Namely,  a setup lacking time-reversal symmetry and for which   ${\cal M}_{12}/{\cal M}_{21}<0$  is always far away from the strong-coupling limit.

Adding a notation for the deviation of ${\cal M}$ from complete symmetry
\begin{align}
\gamma^{}_{a}=\sqrt{|{\cal M}^{}_{12}/{\cal M}^{}_{21}|}\ , 
\end{align}
the positiveness of the Onsager matrix, $
{\cal M}^{}_{11}{\cal M}^{}_{22}-({\cal M}^{}_{12}+{\cal M}^{}_{21})^{2}/4\geq 0$, imposes a relation between $\zeta$ and the asymmetry parameter $\gamma_{a}$
\begin{align}
\sqrt{\Big |\frac{1-\zeta}{1+\zeta}\Big |}\leq\gamma^{}_{a}\leq\sqrt{\Big |\frac{1+\zeta}{1-\zeta}\Big |}\ .\label{conda1}
\end{align}
[For a time-reversal invariant system, $\gamma_{a}=1$ and  condition (\ref{conda1}) is always satisfied.]
With this parametrization of the Onsager matrix, to which we add for convenience
\begin{align}
\gamma=\sqrt{{\cal M}^{}_{11}/{\cal M}^{}_{22}}\ ,
\end{align}
we obtain for the cooling power
\begin{align}
P=\frac{\eta^{-1}_{\rm C}{\cal M}^{}_{21}}{T^{}_{R}}\Big ((\eta^{}_{\rm C}V)-s
\frac{\gamma^{}_{a}/\gamma}{\sqrt{|1-\zeta^{2}|}}
\Big )\ ,\label{P}
\end{align}
where
$s\equiv {\rm sgn}[{\cal M}^{}_{21}]$,
and for the  used power
\begin{align}
W=\frac{\eta^{-2}_{\rm C}{\cal M}^{}_{11}}{T^{}_{R}}\Big ((\eta^{}_{\rm C}V)[(\eta^{}_{\rm C}V)-\frac{\gamma^{}_{a}}{\gamma}
\sqrt{|1-\zeta^{2}|}]\Big )\ . \label{W}
\end{align}
Hence, when ${\cal M}_{21}$ is positive, 
both used and cooling powers are positive for
\begin{align}
(\eta^{}_{\rm C}V)\geq
(\gamma^{}_{a}/\gamma)/\sqrt{1-\zeta^{2}}\ , \label{up}
\end{align}
while if it is negative then the voltage drop $V$ must be negative as well, with
\begin{align}
-(\eta^{}_{\rm C}V)\geq (\gamma^{}_{a}/\gamma)/\sqrt{\zeta^{2}-1}\ .
\end{align}
Finally, the efficiency is
\begin{align}
\frac{\eta}{\eta^{}_{\rm C}}=s\frac{\sqrt{|1-\zeta^{2}|}}{\gamma\gamma^{}_{a}}\frac{(V\eta^{}_{\rm C})-s(\gamma^{}_{a}/\gamma)/\sqrt{|1-\zeta^{2}|}}{
(\eta^{}_{\rm C}V)[(\eta^{}_{\rm C}V)-(\gamma^{}_{a}/\gamma)\sqrt{|1-\zeta^{2}|}]}\ .\label{eff}
\end{align}

\begin{figure}[htp]
\centering
\includegraphics[width=6cm]{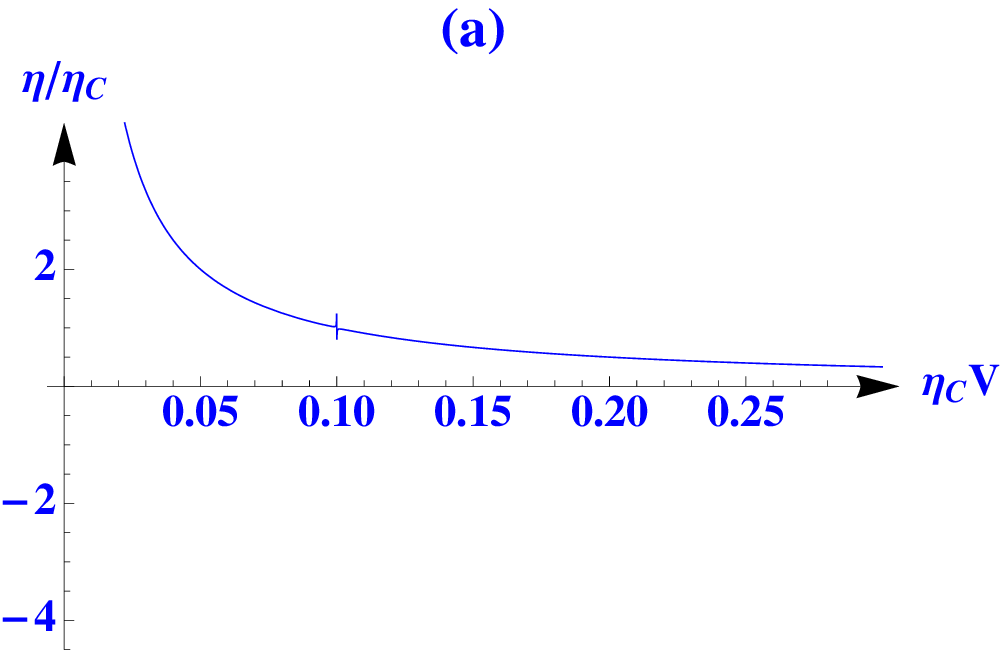}
\includegraphics[width=6cm]{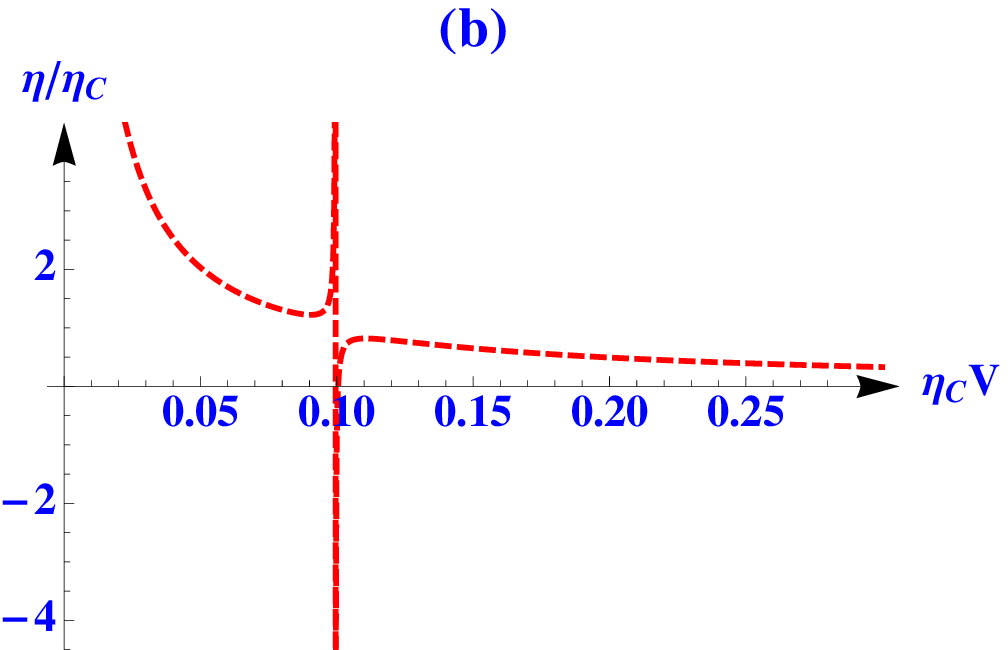}
\includegraphics[width=6cm]{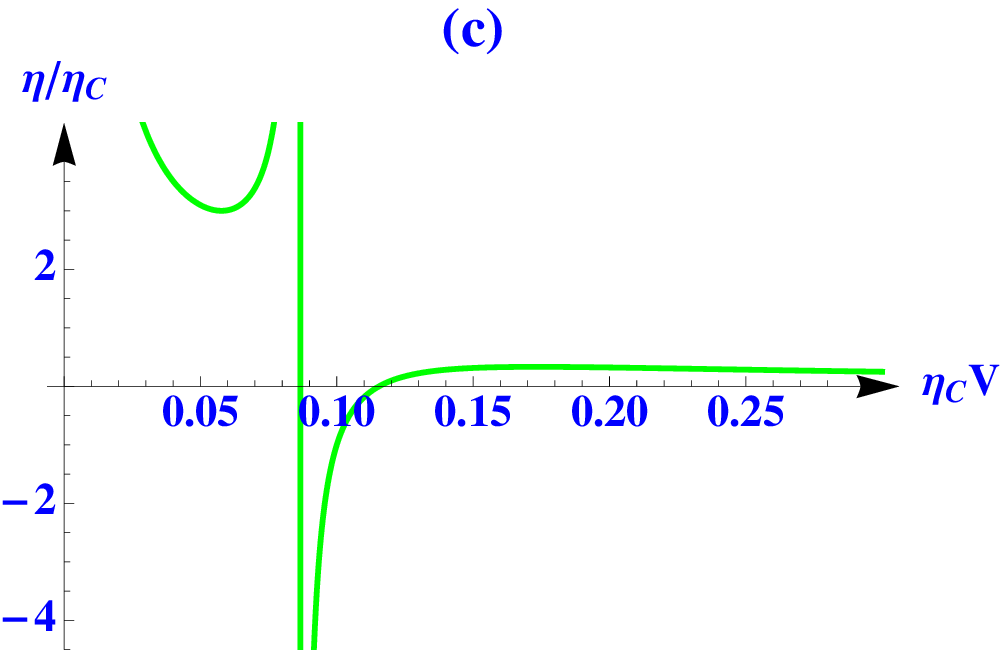}
\caption{The mathematical function of the efficiency, Eq. (\ref{eff}), as a function of $V \eta^{}_{\rm C}$,  for $\gamma_{a}=1$, $\gamma =10$,  and $s=1$, for various values of $\zeta$, (a)-$\zeta =0.01$, (b)-$\zeta=0.1$, and (c)-$\zeta=0.5$.
}
\label{figA}
\end{figure}

The  efficiency is optimized for a given $\eta^{}_{\rm C}$, by controlling $V\eta^{}_{\rm C}$, the ratio of the two components of the driving-force  vector. To do that,
we  first show in Fig. \ref{figA} graphs of the efficiency $\eta / \eta^{}_{\rm C}$ as a  function of $V\eta^{}_{\rm C}$ in the whole range, and with $\gamma_a = 1$, for which the {\em physical} bound on  $V\eta^{}_{\rm C}$ is given in Eq. (\ref{up}). The mathematical function  of the efficiency has a   ``pole" at
$V\eta^{}_{\rm C} = \sqrt{1-\zeta^{2}}/\gamma$ and a zero at $V\eta^{}_{\rm C}= (\gamma \sqrt{1-\zeta^{2}})^{-1}.$
It is large in the non-physical region on the left of the pole and negative between the pole and the zero. We emphasize that although the efficiency can be larger in the non-physical regions,  we have to choose the maximum within the physical domain, given by Eq.  (\ref{up}). The plots in Fig. \ref{figA} help us to do that. The physical region is on the right of both zero and pole.

When the system tends to the    ``ideal" limit $\zeta \rightarrow 0$, the zero and the pole merge together and the maximum of the efficiency, $\eta/\eta^{}_{\rm C} \rightarrow  1$ is infinitesimally above $V \eta^{}_{\rm C} = 1/\gamma$.
This rather nontrivial behavior is clearly seen in Fig. \ref{figA}a, and Eqs.  (\ref{P}) and (\ref{W})   show how both the cooling and  used powers approach zero at the ideal point. The pole and the zero get further apart as $\zeta$ increases, Figs. \ref{figA}b and \ref{figA}c, and 
the value of
$V \eta^{}_{\rm C}$ for which the efficiency is maximal increases as well (see below).

The efficiency is optimal for a given $\eta^{}_{\rm C}$ at $V=V_{\rm m}$, where
\begin{align}
V^{}_{\rm m}\eta^{}_{\rm C}=s\frac{\gamma^{}_{a}}{\gamma}\sqrt{\Big |\frac{1+\zeta}{1-\zeta}\Big |}\ ,\label{choice}
\end{align}
showing that the sign of the bias is determined by the  $s\equiv{\rm sgn}{\cal M}_{21}$, as expected.
At the voltage drop corresponding to the optimal efficiency, the cooling power is 
\begin{align}
P^{}_{\rm m}=\frac{\eta^{-1}_{\rm C}|{\cal M}^{}_{21}|}{T^{}_{R}}\frac{\gamma^{}_{a}}{\gamma}\frac{\zeta}{\sqrt{|1-\zeta^{2}|}}\ ,
\end{align}
the  used power is
\begin{align}
W^{}_{\rm m}=\frac{\eta^{-2}_{\rm C}{\cal M}^{}_{11}}{T^{}_{R}}\frac{\gamma^{2}_{a}}{\gamma^{2}_{}}\zeta\Big |\frac{1+\zeta}{1-\zeta}\Big |\ , 
\end{align}
and the optimal efficiency $\eta_{\rm m}$ is 
\begin{align}
\eta^{}_{\rm m}=\eta^{}_{\rm C}\frac{1}{\gamma^{2}_{a}}\Big |\frac{1-\zeta}{1+\zeta}\Big |\ .\label{opt}
\end{align}
Note that the optimal efficiency never exceeds the Carnot one: this is ensured by the condition (\ref{conda1}).
Finally, 
the total dissipation at optimal efficiency is
\begin{align}
T^{}_{R}\dot{S}^{}_{\rm m}
&=\Bigl (W^{}_{\rm m}-\eta^{-1}_{\rm C}P^{}_{\rm m}
\Bigr )\nonumber\\
&=\eta^{-2}_{\rm C}{\cal M}^{}_{22}\frac{\zeta}{|1-\zeta|}\Bigl (\gamma^{2}_{a}(1+\zeta)-(|1-\zeta |)\Bigr )\ .
\end{align}
Hence we see that as a time-reversal symmetric junction ($\gamma_{a}=1$) approaches the strong-coupling limit, 
its efficiency, cooling power, and the used power all deviate from the ideal limit as $\zeta$;  the total dissipation, however, deviates much more weakly from zero, as $\zeta^{2}$, due to the cancellation of the $\zeta$ term between the thermal and the electrical dissipations.

\begin{figure}[htp]
\includegraphics[width=6cm]{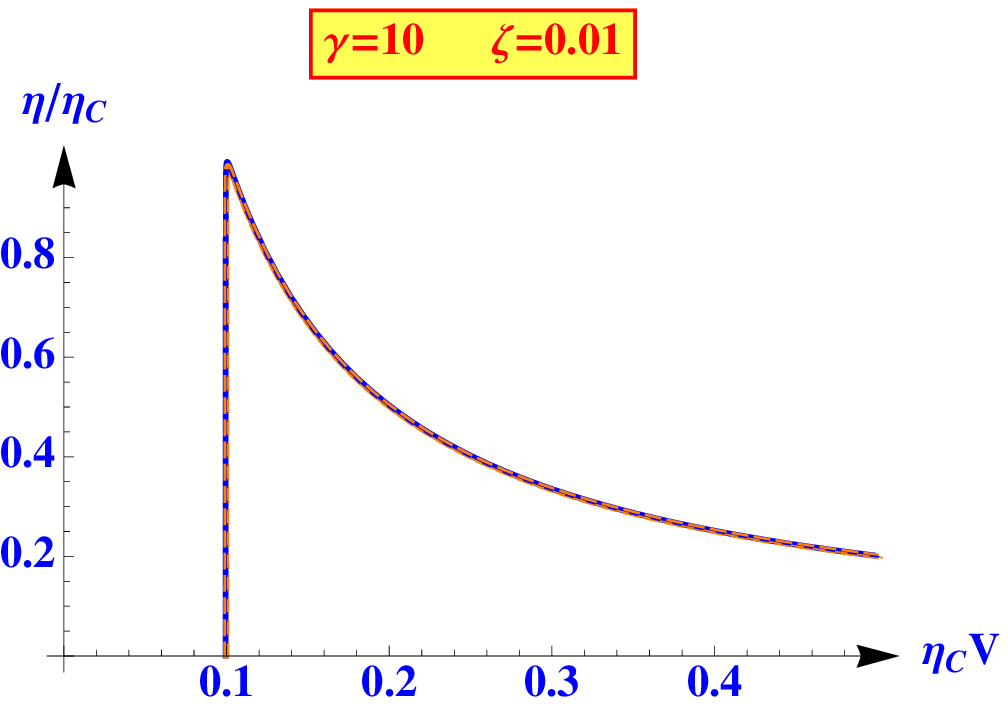}
\includegraphics[width=6cm]{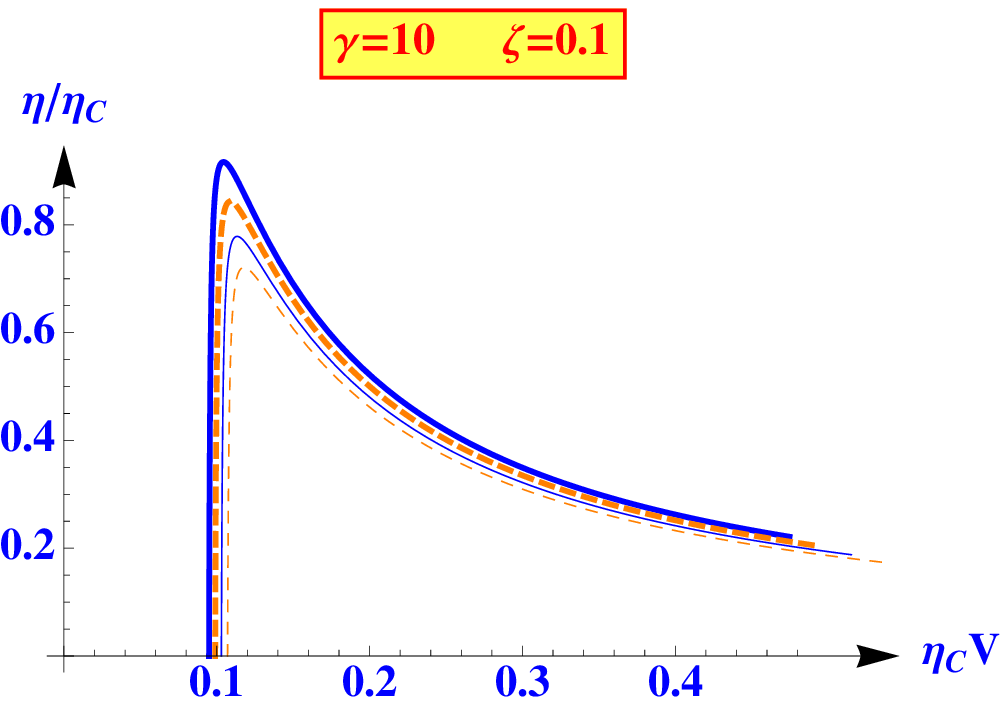}
\includegraphics[width=6cm]{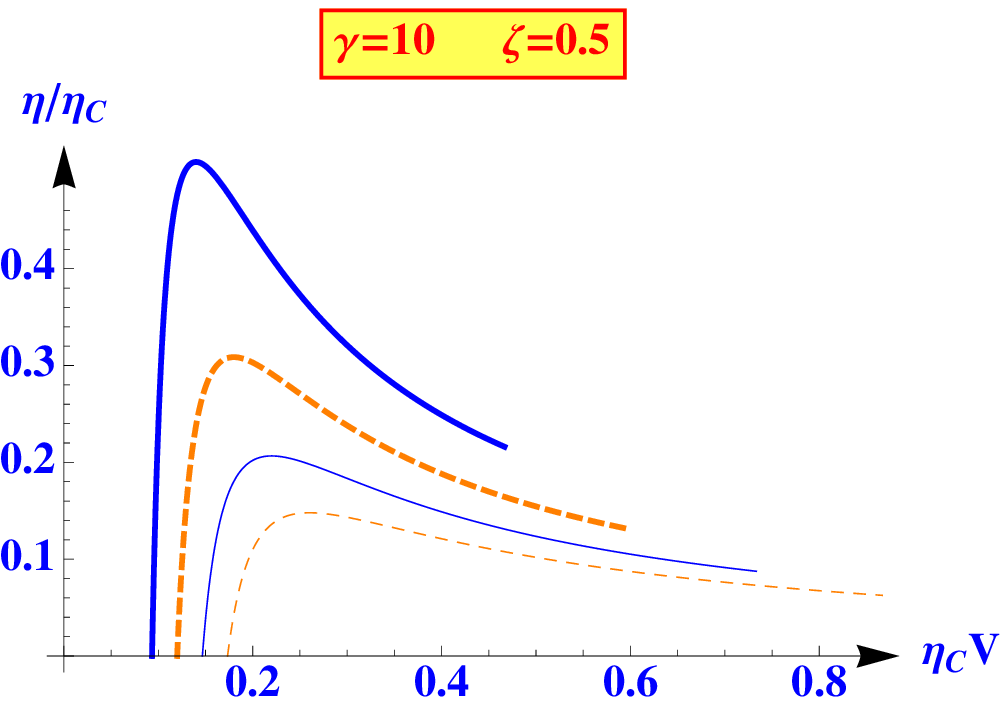}
\caption{The efficiency as a function of the driving force, in the physical region. The values of $\zeta$ and $\gamma$ are marked on each panel; the four curves, in decreasing order, correspond to different values of the asymmetry parameter, in increasing order according to Eq. (\ref{conda1}); $\gamma_{a}=0.994,0.998,1.002,1.006$ for the upper panel, $\gamma_{a}=0.945, 0.985, 1.025, 1.065$ for the  next panel, and $\gamma_{a}=0.81, 1.04, 1.27, 1.5$ for the lower panel.
}
\label{fig2}
\end{figure}

The effect of time-reversal symmetry breaking (for a positive ratio of the two off-diagonal elements of the matrix ${\cal M}$)
is examined in Figs. \ref{fig2}. The figures show the dependence of the efficiency on the driving force in the physical region of the latter [see Eq. (\ref{up})]. The four curves in each panel are for various values of the asymmetry parameter $\gamma_{a}$, according to  condition (\ref{conda1}), with the efficiency   decreasing as $\gamma_{a}$  is increasing [see Eq. (\ref{opt})], {\em for the parameters used here.}
These show that  increasing $\gamma_{a}$ is detrimental to the coefficient of performance of cooling, the more so as the deviation from the strong-coupling limit increases. This asymmetry has another outcome. Whereas for $\gamma_{a}=1$ a subtle cancellation of the order $\zeta-$terms in the dissipation causes it to deviate less from the ideal limit as compared to the powers, this {\em ceases to hold as} $\gamma_{a}\neq 1$. This is a notable result of our analysis.

\subsection{Analysis of the eigenvalues and eigenvectors of the Onsager matrix with time-reversal symmetry, ($\gamma_{a}=1$).} \label{vec}

It is interesting to study the behavior of the driving-force vector close to the ideal limit, where the dissipation is very small. \cite{COM4}
As the only variable is $\eta_{\rm C}V$, 
it suffices to consider just the ratio of the first to the second component of the eigenvectors.
Writing the Onsager matrix for $\gamma_{a}=1$  in the parametrized form (for brevity, we take both off-diagonal elements of ${\cal M}$ to be positive)
\begin{align}
{\cal M}\Rightarrow\left [\begin{array}{cc}\gamma &\sqrt{1-\zeta^{2}} \\   &  \\ \sqrt{1-\zeta^{2}}&1/\gamma\end{array}\right ]\ ,
\end{align}
we obtain that in the small $\zeta-$limit, the lower and the higher eigenvalues, $\lambda_{\pm}$, are
\begin{align}
&\lim_{\zeta\rightarrow 0}\lambda^{}_{-}\rightarrow \zeta^{2}/(\gamma+\gamma^{-1})\ ,\nonumber\\
&\lim_{\zeta\rightarrow 0}\lambda^{}_{+}\rightarrow
\gamma+\gamma^{-1}_{}-
\zeta^{2}/(\gamma+\gamma^{-1})\ .
\end{align}
The corresponding (unnormalized) eigenvectors, $v_{\pm}$, are
\begin{align}
&\lim_{\zeta\rightarrow 0}v^{}_{-}\rightarrow\Big \{-\gamma^{-1}\Big (1-\frac{\zeta^{2}}{2}\frac{\gamma^{2}-1}{\gamma^{2}+1}\Big ),1\Big \}\ ,\nonumber\\
&\lim_{\zeta\rightarrow 0}v^{}_{+}\rightarrow\Big \{\gamma^{}\Big (1+\frac{\zeta^{2}}{2}\frac{\gamma^{2}-1}{\gamma^{2}+1}\Big ),1\Big \}\ .
\end{align}
We have found that  at optimal  efficiency 
our driving-force  vector is [see Eq. (\ref{choice})]
\begin{align}
v^{}_{\rm m}\propto
\Big \{\frac{1}{\gamma}\sqrt{\frac{1+\zeta}{1-\zeta}},-1\Big \}\ .
\end{align}
Hence $v_{\rm m}$ is directed along the eigenvector belonging to the lower eigenvalue at $\zeta\rightarrow 0$, and rotates away from it as the Onsager matrix deviates from the ideal  strong-coupling limit.

\section{Discussion of the asymmetry of the dissipation {\it  vs.} the thermopower}
 \label{dissa}

The treatment of the heat exchange between transport electrons and an electron reservoir is based on the principle that a particle with an energy $E$ exiting (entering) a large system extracts from (delivers to) it a heat $\Delta Q$ and an entropy $\Delta S$ given by 
\begin{align}
\Delta Q = (E - \mu); ~~~~\Delta S = (E - \mu) / T\ , \label{basic}
\end{align}
where $\mu$ is the chemical potential in the reservoir.
This is proven thermodynamically, for example, in Ref.~ \onlinecite{Ziman}, adopted to the mesoscopic regime within the Landauer-transport-formulation in Ref.~ \onlinecite{SIVAN},  and given a statistical-mechanics proof in Ref. ~\onlinecite{Ariel}. For two reservoirs with chemical potentials $\mu_L$ and 
$\mu_R$, connected by a single-channel   ``wire" 
having a transmission coefficient ${\cal T}(E)$, this yields \cite{SIVAN}  for the heat $-J^{Q}_{\alpha}\equiv\Delta Q^{}_{\alpha} = T \Delta S^{}_{\alpha}$   delivered to reservoir $\alpha $   ($\alpha =L$ or $R$)
\begin{align}
-J^{Q}_{\alpha }= \frac{2}{h}\int_{-\infty}^{\infty} (\mu^{}_{\alpha} - E){\cal T}(E) [f^{}_{\alpha}(E) - f^{}_{\beta}(E)] dE\ ,
\label{heat}
\end{align} 
with $\beta \ne \alpha$ and $f_{L,R}(E)$ being the corresponding Fermi function.
We limit ourselves to the linear-response regime with a single driving field $V = (\mu_L -\mu_R)/e$. The electrical current between the reservoirs gives the conductance $G$ and the heat current gives the Onsager coefficient ${\cal L}\equiv{\cal M}^{}_{21}/T$, [cf. Eq. (\ref{onsager})]. The Seebeck thermopower ${\cal S}$ is then given by ${\cal S} ={\cal  L}/(GT)$. We consider here only the case where time-reversal symmetry prevails.

The first crucial observation here from Eq. (\ref{heat}) is the well-known one that in order to have nonzero thermoelectric response, ${\cal T}(E)$
must have an odd component as a function of $E - \mu$. Next, we note that if such ``electron-hole ($e-h$) symmetry breaking" exists, the two powers $J^{Q}_{L,R}$ are each linear in $V$. However, by adding them, one finds that the total dissipation (heat generated  per unit time) is indeed (see Ref.~ \onlinecite{SIVAN}) the good old Joule heating, $IV$, which is second-order in $V$ ($I$ is the net electrical current).

The novel observation of Ref. ~\onlinecite{JCC} is that it is the same symmetry breaking which also leads to an {\em asymmetry of the dissipation between the two reservoirs}. In fact, by subtracting Eq. (\ref{heat}) for $L$ and $R$,
one readily obtains, to order $V$,
\begin{align}
J^{Q}_R(V) -J^{Q}_{L}(V) = 2{\cal L}V =  2 G{\cal S}TV\ .
\end{align}
Thus the dissipation asymmetry and the thermopower are proportional to one another, at least within linear response. The physical reason for this is clear: within linear response, the energy flow from, say, $L$ to $R$, which differs by just a constant from the heat flow, can be regarded as less energy (and heat) dumped in $L$,  and more dumped in $R$. This transfers energy from $L$ to $R$.

An interesting feature is that the heat flow/dissipation asymmetry, while small for weak $e-h$ symmetry breaking, is first order in V. The total dissipation, which exists also with that symmetry valid, is second order--i.e. much smaller!
As shown in Ref. ~\onlinecite{JCC}, this thermopower--dissipation-asymmetry relationship can be used to determine the former quantity by measuring the latter. We note that when the Seebeck coefficient,  ${\cal S} = 0$,  vanishes  
the dissipation is symmetric and independent of the nature of the two reservoirs (and of any asymmetry between them).

Although in linear response the currents are odd in $V$, the dissipation {\em in each reservoir} is not. 
In fact, by subtracting Eq. (\ref{heat}) for $L$, with $V$ and $-V$, it is found \cite{JCC} that, e,g. 
\begin{align}
J^{Q}_L(-V) -J^{Q}_L(V) = 2{\cal L}V =  2 G{\cal S}TV\ ,
\end{align}
as well.
These relationships should hold for any meso- and nano- scale system, including molecular bridges.
They were confirmed by both electronic-structure numerical computations and by experiments using an innovative nano-scale temperature detection.\cite{JCC} 

We conclude 
by discussing some simple particular cases, demonstrating the  relevance to our main subject in the present paper--the case of small dissipation. Having full $e-h$ symmetry is of no interest here, since both ${\cal S}$ and the dissipation asymmetry vanish. A very interesting case is that a almost full $e-h$ symmetry breaking, when almost all charge carriers are e.g. electrons. A simple realisation of this is a narrow Lorentzian resonance of width $\Gamma$ much above the Fermi level (i.e when its center energy $E_0$ satisfies $E_0 - \mu >> \Gamma, k_BT$). In the low-temperature and narrow-resonance limit this becomes the strong-coupling configuration \cite{Kedem} or `best'   \cite{Mahan} thermoelectric, having a singular transport matrix. The system then has vanishing dissipation and power. 
The new insight gained from the thermopower--dissipation-asymmetry relationship is that the dissipation is negative in one terminal and positive in the other, such that they fully cancel each other (while heat is still carried by the electron current from $L$ to $R$).

A finite width of the resonance, or a low but significant temperature will create the almost singular transport matrix discussed in this paper. A more complicated case is that of a mobility edge far above the Fermi level. Again, most charge carriers will be electrons. This case was treated in Refs. ~\onlinecite{SIVAN} and ~\onlinecite{Ariel}, and sizeable values of the Seebeck coefficient  ${\cal S}$ were indeed found. This entails large dissipation asymmetry.

\section*{Acknowledgments}

This work was supported by the Israeli Science Foundation (ISF) and the US-Israel Binational Science Foundation (BSF).


\begin{thebibliography}{99}
 

\bibitem{book} 
T. C. Harman and J. M. Honig, {\it Thermoelectric and
Thermomagnetic Effects and Applications} (McGraw-Hill, New-York,
1967); H. Julian Goldsmid,  {\it Introduction to Thermoelectricity}, Springer Series in Material Science (2009).

 
\bibitem{Rutten}
B. Rutten, M. Esposito, and B. Cleuren, Phys. Rev. B {\bf 80}, 235122 (2009).
 
 
 \bibitem{we}
O. Entin-Wohlman, Y. Imry, and A. Aharony, Phys. Rev. B {\bf 82}, 115314 (2010).


 \bibitem{Cleuren}
 
 
B. Cleuren, B. Rutten, and C. Van den Broeck, Phys. Rev. Lett. {\bf 108}, 120603 (2012). 
 
 
 


 
\bibitem{JHJPAP}
J-H. Jiang, O. Entin-Wohlman, and Y. Imry, Phys. Rev. B {\bf 85}, 075412 (2012); {\it ibid.}
B {\bf 87}, 205420 (2013).

\bibitem{NJP} 
J.-H. Jiang, O. Entin-Wohlman,  and Y.  Imry, New J. Phys. {\bf 15}, 075021 (2013).
















\bibitem{Kedem}

O. Kedem and S. R. Caplan, Trans. Faraday Soc. {\bf 61}, 1897 (1965); see also 
W. Shockley and H. J. Queisser, J. Appl. Phys. {\bf 32}, 510 (1961).

\bibitem{Mahan}
G. D.  Mahan and J. O.  Sofo,  Proc. Natl. Acad., USA {\bf 93},  7436 (1996).




\bibitem{gordon} 

See J. M. Gordon, Am. J. Phys. {\bf 59}, 551 (1991) and references therein.






\bibitem{COM2}
The condition adopted in Ref. ~ \onlinecite{Cleuren} is $\mu_{L}=\mu_{R}$.  When this is the case,  the sum of the heat currents vanishes as well. 



\bibitem{mybook}
Y. Imry, {\it Introduction to Mesoscopic Physics} 2nd ed., Oxford University Press, New York 2002.





\bibitem{JCC} 
W. Lee,	 K. Kim,	 W. Jeong,	 L. A. Zotti,	 F. Pauly,	 J. C. Cuevas,	 and P. Reddy, Nature {\bf 498}, 209 (2013);
L. A. Zotti, 
M. B\"{u}rkle, F. Pauly, W. Lee, K. Kim,  
W. Jeong, Y. Asai, P. Reddy, 
and J. C. Cuevas, arXiv:1307.8336.


\bibitem{Benenti}

G. Benenti, K. Saito, and G. Casati, Phys. Rev. Lett. {\bf 106}, 230602 (2011).




\bibitem{Seifert}

K. Brandner, K. Saito, and U. Seifert, Phys. Rev. Lett. {\bf 110}, 070603 (2013); K. Brandner and U. Seifert, New J. Phys. (in press) arXiv:
1308.2179.



\bibitem{COM3}
A minuscule determinant of the Onsager matrix ${\cal M}$ corresponds to a very large figure of merit.


\bibitem{COM4}
We consider in Sec. \ref{vec} only the symmetric case $\gamma_{a}=1$ as in the asymmetric one there is a distinction between the left and right eigenvectors.










\bibitem{Ziman}
J. M. Ziman, {\it Principles of the Theory of Solids}, Cambridge University Press (1969); see sections 7.7-7.9.

\bibitem{SIVAN}
U. Sivan and Y. Imry, 
Phys. Rev. B {\bf 33}, 551 (1986).

\bibitem{Ariel} 
Y. Imry and Y. Amir,        
in: {\it 50 Years of Anderson Localization}, E. 
Abrahams, ed., 
World Scientific 2010.  





\end{thebibliography}
\end{document}